\documentclass[conference]{IEEEtran}
\usepackage{cite}
\usepackage{amsmath,amssymb}
\usepackage{graphicx}
\usepackage{booktabs}
\usepackage{xcolor}
\usepackage[hidelinks]{hyperref}

\newcommand{\CAD}{$-0.0136$}
\newcommand{\CALo}{$-0.0178$}
\newcommand{\CAHi}{$-0.0098$}

\newcommand{\CAPR}{$-0.027$}
\newcommand{\CAF}{$-0.024$}
\newcommand{\CBD}{$-0.0056$}
\newcommand{\CBLo}{$-0.0095$}
\newcommand{\CBHi}{$-0.0021$}

\newcommand{\CCD}{$+0.0013$}
\newcommand{\CCLo}{$-0.0003$}
\newcommand{\CCHi}{$+0.0031$}

\newcommand{\CDD}{$+0.0228$}
\newcommand{\CDLo}{$+0.0187$}
\newcommand{\CDHi}{$+0.0265$}

\newcommand{\CED}{$+0.0064$}
\newcommand{\CELo}{$+0.0034$}
\newcommand{\CEHi}{$+0.0094$}

\newcommand{\CGD}{$+0.0166$}
\newcommand{\CGLo}{$+0.0118$}
\newcommand{\CGHi}{$+0.0208$}

\newcommand{\CHD}{$-0.0109$}
\newcommand{\CHLo}{$-0.0142$}
\newcommand{\CHHi}{$-0.0076$}

\newcommand{\CID}{$-0.0022$}
\newcommand{\CILo}{$-0.0046$}
\newcommand{\CIHi}{$+0.0003$}

\newcommand{\NTest}{25,596}
\newcommand{\NTestPat}{2,797}
\newcommand{\NVal}{8,536}
\newcommand{\NTrain}{77,988}
\newcommand{\NTrainPat}{25,207}
\newcommand{\NValPat}{2,801}
\newcommand{\EvaxAUROC}{0.828}
\newcommand{\EvaxSharedAUROC}{0.798}
\newcommand{\DenseAUROC}{0.808}
\newcommand{\DenseEvaxRecAUROC}{0.805}
\newcommand{\VitAUROC}{0.705}
\newcommand{\VitFixAUROC}{0.787}
\newcommand{\FiveAvgAUROC}{0.814}
\newcommand{\SixAvgAUROC}{0.822}

\newcommand{\FiveAvgECE}{0.397}
\newcommand{\FiveCalAvgECE}{0.013}

\newcommand{\SixAvgECE}{0.330}

\newcommand{\FiveCost}{1.81}
\newcommand{\EvaxCost}{0.23}
\newcommand{\CostRatio}{7.8}
\newcommand{\FiveParams}{217}
\newcommand{\EvaxParams}{22}

\newcommand{\PubEvax}{0.833}
\newcommand{\SubFMA}{0.828}

\newcommand{\SubFMC}{0.829}
\newcommand{\SubFMD}{0.827}
\newcommand{\SubFME}{0.826}
\newcommand{\SubFMF}{0.823}
\newcommand{\SubFMBestSize}{2}
\newcommand{\SubFMBestAUROC}{0.831}

\newcommand{\SubFiveE}{0.813}
\newcommand{\SubFiveBestSize}{2}
\newcommand{\SubFiveBestAUROC}{0.815}
\newcommand{\PLNeg}{11}

\newcommand{\PLSigNegNames}{Atelectasis, Effusion, Pneumothorax, Pleural Thickening}

\newcommand{\AblNoAug}{$-$0.025}
\newcommand{\AblNoFocal}{$-$0.007}
\newcommand{\LWCalMin}{+0.0013}
\newcommand{\LWCalMax}{+0.0052}
\newcommand{\StackCalMinAll}{$-0.0324$}
\newcommand{\StackCalMaxAll}{$-0.0015$}
\newcommand{\NComp}{29}

\renewcommand{\PLSigNegNames}{Atelectasis, Effusion, Pleural Thickening}

\newcommand{\SeedEvaxMin}{0.827}
\newcommand{\SeedEvaxMax}{0.828}

\newcommand{\SeedEvaxSD}{0.0006}
\newcommand{\SeedDenseMin}{0.805}
\newcommand{\SeedDenseMax}{0.809}

\newcommand{\SeedDenseSD}{0.0022}
\newcommand{\SeedGapMin}{+0.0186}
\newcommand{\SeedGapMax}{+0.0236}
\newcommand{\SeedFiveGapMin}{$-$0.0139}
\newcommand{\SeedFiveGapMax}{$-$0.0128}
\newcommand{\FiveFullAUROC}{0.817}

\newcommand{\VitEvaxRecAUROC}{0.785}
\newcommand{\VitFixFiveAUROC}{0.817}
\newcommand{\VitFixFiveGap}{$-$0.0110}

\begin{document}

\title{Is a CNN--Transformer Ensemble Worth It When a\\Chest X-Ray Foundation Model Is Available?}

\author{
\IEEEauthorblockN{Aditya Seelaboyina}
\IEEEauthorblockA{Department of Computer Science and Engineering\\
Texas A\&M University\\
College Station, TX, USA\\
aditya\_0416@tamu.edu}
\and
\IEEEauthorblockN{Irfan Khan}
\IEEEauthorblockA{Department of Marine Engineering Technology,\\
in a courtesy joint appointment with the\\
Department of Electrical \& Computer Engineering\\
Texas A\&M University, TX, USA\\
irfankhan@tamu.edu}
}

\maketitle

\begin{abstract}
Ensembles of ImageNet-pretrained convolutional networks and Vision Transformers are a common way to improve multi-label chest X-ray classification. We ask whether such an ensemble is still worth its cost when a model pretrained on chest radiographs is available. On the official patient-disjoint split of NIH ChestX-ray14 (\NTest{} test images, \NTestPat{} patients), we train five ImageNet-pretrained backbones and the EVA-X-S foundation model with validation-only model selection, compare five fusion rules, and estimate every difference with a patient-level paired bootstrap. EVA-X-S and DenseNet121 were trained with three seeds. EVA-X-S alone reached a test macro-AUROC of \EvaxAUROC; the five-model ensemble reached \FiveAvgAUROC{} (difference \CAD, 95\% CI \CALo{} to \CAHi) at about \CostRatio{} times the summed backbone latency and ten times the parameters. Averaging the five ImageNet models with EVA-X-S changed AUROC by \CBD{} (\CBLo{} to \CBHi). Trained with the same recipe as EVA-X-S (apart from layer-wise learning-rate decay), the strongest ImageNet model (DenseNet121) scored \DenseEvaxRecAUROC, and with all five ImageNet backbones trained with that recipe their ensemble scored \FiveFullAUROC, still \CHD{} (\CHLo{} to \CHHi) below EVA-X-S. Across seeds, EVA-X-S varied by at most 0.001 AUROC. Under the ImageNet models' recipe, however, EVA-X-S fell to \EvaxSharedAUROC. Our full-stacking configuration never outperformed calibrated averaging, and its calibration benefit was reproduced by per-model recalibration. In this setting, the tested ImageNet ensembles did not improve on a single chest X-ray foundation model fine-tuned with its own recipe.
\end{abstract}


\section{Introduction}
Chest radiography is the most commonly performed diagnostic imaging examination~\cite{raoof2012}, and multi-label classification of its findings is a standard benchmark task~\cite{wang2017}. A frequent way to improve such classifiers is to combine several architectures, typically convolutional neural networks (CNNs) and Vision Transformers, by averaging their outputs or by training a meta-classifier on them~\cite{ashraf2023synth,li2024chexds,hsieh2026hybrid}. The rationale is that architectures with different inductive biases make partly independent errors~\cite{dietterich2000,zaidi2021nes}.

Chest X-ray foundation models pretrained on hundreds of thousands of radiographs, such as EVA-X~\cite{yao2025evax} and Ark+~\cite{ma2025ark}, now report single-model results on ChestX-ray14 that match or exceed published ensembles. Among the ensemble studies we reviewed, none includes such a model as a baseline or a member, none reports calibration or measured inference cost, and most evaluate on custom splits without confidence intervals (Section~\ref{sec:related}). Foundation-model papers compare against single ImageNet-initialized baselines~\cite{yao2025evax,ma2025ark,moutakanni2024raydino}, and recent benchmarks compare foundation models with single CNNs~\cite{inacio2026adults}. What remains untested is the marginal value of a heterogeneous ImageNet ensemble, and its cost, once such a model is available.

We ask: \emph{when a chest X-ray foundation model is available, does a heterogeneous ensemble of ImageNet-pretrained CNNs and Transformers still improve multi-label classification enough to justify its cost?} We answer under one protocol on the official patient-disjoint split, with checkpoints, fusion weights, calibrators, thresholds, and the choice of fusion rule all selected on validation data.

Our contributions are:
\begin{enumerate}
\item A comparison on the official split showing that a 22M-parameter chest X-ray foundation model outperforms a 217M-parameter five-backbone ImageNet ensemble, and that averaging the ensemble's members with it does not help.
\item A recipe control with seed replicates: none of the ImageNet backbones, alone or as an ensemble, closes the gap when trained like the foundation model, while the foundation model itself loses its advantage under the ensemble's training recipe.
\item A fusion analysis showing that our full-stacking configuration does not outperform calibrated averaging, and that its calibration benefit comes from recalibration.
\end{enumerate}

\section{Related Work}\label{sec:related}

\subsection{Ensembles for Chest X-Ray Classification}
SynthEnsemble~\cite{ashraf2023synth} combines six CNN, Transformer, and hybrid networks on ChestX-ray14 and finds that a validation-weighted average (mean AUROC 0.854) outperforms both the best single model (0.842) and XGBoost stacking, on a custom split without confidence intervals. CheX-DS~\cite{li2024chexds} combines DenseNet121 and Swin-B by weighted averaging (0.838 versus 0.828). Hsieh \textit{et al.}~\cite{hsieh2026hybrid} compare voting, weighted fusion, and logistic-regression stacking of CNN--Transformer hybrids on a patient-level split with bootstrap intervals and external tests, reporting internal gains of about 0.01 AUROC over DenseNet121. None of these studies includes a chest X-ray foundation model, measures calibration, or reports inference latency, and their splits differ from the official one.

\subsection{Chest X-Ray Foundation Models}
EVA-X~\cite{yao2025evax} pretrains Vision Transformers by masked image modeling on about 520k unlabeled frontal radiographs from ChestX-ray14, CheXpert, and MIMIC-CXR, with test images excluded according to the authors, and reports official-split mean AUROC of 0.824--0.835. Its released fine-tuning script uses no validation set and keeps the checkpoint with the highest test-set AUROC, so we fine-tune it ourselves with validation-based selection. Ark+~\cite{ma2025ark} accrues knowledge from many expert-labeled datasets, including ChestX-ray14 in the earlier Ark work~\cite{ma2023ark}. RayDINO~\cite{moutakanni2024raydino}, a self-supervised X-ray foundation model, reports 0.838 AUROC on ChestX-ray14. The CXR-LT 2026 challenge reports that vision-language foundation models improve long-tailed classification~\cite{cxrlt2026}, and benchmarks of RAD-DINO, BiomedCLIP, and other encoders against single CNNs find that rankings depend on the adaptation strategy and on domain shift~\cite{inacio2026adults}. Like EVA-X, these works compare against single models, not ensembles.

\subsection{Ensembles and Calibration}
Deep ensembles improve accuracy and predictive uncertainty~\cite{lakshminarayanan2017}, and architecturally diverse ensembles can be more robust under distribution shift~\cite{zaidi2021nes}. Stacked generalization~\cite{wolpert1992} learns how to weight members, and greedy ensemble selection~\cite{caruana2004} chooses which to keep. Modern networks are often miscalibrated~\cite{guo2017}, and class-reweighted training shifts probabilities further, so part of a learned fuser's benefit may be recalibration.

\section{Methods}\label{sec:method}

\subsection{Data}
We use NIH ChestX-ray14~\cite{wang2017} (112,120 frontal radiographs, 30,805 patients, 14 report-mined findings). The test set is the official test list (\NTest{} images, \NTestPat{} patients). From the official training list we hold out 10\% of patients for validation (\NVal{} images, \NValPat{} patients) and train on the remainder (\NTrain{} images, \NTrainPat{} patients). No patient appears in more than one split (Table~\ref{tab:splits}). Images are converted to 8-bit grayscale, resized to $224\times224$ by bilinear interpolation without cropping, replicated to three channels, and normalized with ImageNet channel statistics for every model, including EVA-X-S.

\subsection{Backbones and Training Recipes}
ResNet18~\cite{resnet}, DenseNet121~\cite{densenet}, CoAtNet-0~\cite{coatnet}, ViT-B/16~\cite{vit}, and Swin-B~\cite{swin} start from ImageNet weights in \texttt{timm}~\cite{timm}. EVA-X-S (22M parameters) starts from its public masked-image-modeling weights~\cite{yao2025evax} (\texttt{eva\_x\_small\_patch16\_merged520k\_mim.pt}), loaded with a check that every backbone tensor matches; only the classification head is new. Two recipes are used.

The \emph{shared recipe}, fixed before any results, uses a focal loss ($\gamma=2$)~\cite{focal} with positives weighted by the training negative-to-positive ratio; AdamW (learning rate $10^{-4}$, weight decay $10^{-4}$); batch size 32; a one-epoch warmup and cosine decay over at most 40 epochs; and augmentation by flips, affine transforms, brightness, contrast, gamma, noise, and blur~\cite{albumentations}, plus Mixup ($\alpha=0.4$)~\cite{mixup}. All five ImageNet backbones and one EVA-X-S run use it.

The \emph{EVA-X recipe} follows the EVA-X fine-tuning settings: plain binary cross-entropy, no Mixup, AdamW with weight decay 0.05, drop-path 0.2, layer-wise learning-rate decay 0.55, five warmup epochs, at most 30 epochs, and batch 256 with the learning rate scaled linearly from the original to $2.5\times10^{-4}$. Our augmentation replaces RandAugment, and our ImageNet input normalization differs from the grayscale statistics in the EVA-X code (mean 0.492, s.d. 0.285). EVA-X-S and, as a control, all five ImageNet backbones use this recipe; layer decay applies only to EVA-X-S.

Every run uses seed 42; EVA-X-S and DenseNet121 under the EVA-X recipe were also trained with seeds 1 and 2, which change initialization of the new layers, data order, augmentation, and dropout.

For every run, the checkpoint with the best validation macro-AUROC is kept, with early stopping after 5 epochs without improvement. Training uses bfloat16 on one NVIDIA A100 80GB GPU.

\subsection{Fusion Rules}
Five rules combine member probabilities, all fitted on validation data: probability averaging; calibrated averaging (per-model, per-finding Platt scaling, then averaging logits); weighted averaging (one non-negative weight per model minimizing validation log-loss, then Platt scaling); label-wise stacking (per-finding logistic regression on that finding's member probabilities); and full stacking (per-finding logistic regression on all members' probabilities for all findings). Validation metrics for fitted rules are out-of-fold (5-fold, grouped by patient): each rule is fit on four folds of the members' validation predictions and scored on the fifth, and F1 thresholds are chosen on these out-of-fold predictions. The rule is then refit on all validation data and applied once to the test set. Stacking regressions use L2 regularization ($C=1$, no class weights); Platt scaling is effectively unregularized. The rule used for headline claims is the one with the highest out-of-fold validation macro-AUROC, a choice made before the test set was scored. For each ensemble size, the best subset of members is likewise chosen on validation, using calibrated averaging because it puts members trained with different losses on a common probability scale.

\subsection{Evaluation and Statistics}
The primary endpoint is test macro-AUROC; secondary endpoints are macro-AUPRC, macro-F1 with per-finding thresholds chosen on validation, expected calibration error (ECE, 15 bins, averaged over findings), and summed backbone latency (per-model bfloat16 forward time at batch 64, after warm-up, over three repeats; preprocessing, fusion, and data transfer are excluded). AUROC is computed per image. Differences are estimated with 1,000 paired bootstrap resamples of test patients: patients are drawn with replacement, all of a patient's images are kept, and both systems are scored on the same resample. In each resample every per-finding AUROC is recomputed and macro-averaged over findings with both classes present. We report 95\% percentile intervals and two-sided $p=2(k+1)/(B+1)$, with $k$ the number of resamples on the smaller side of zero, Holm-corrected within each family of comparisons (Table~\ref{tab:comparisons}). This $p$ inverts the percentile interval: it is the smallest level at which the equal-tailed interval excludes zero~\cite{efron1993}, so $p<0.05$ exactly when the 95\% interval excludes zero. It is a bootstrap test of no difference rather than a permutation test, and we did not run a permutation test. This is the primary inference. As an approximate, non-confirmatory sensitivity analysis, Table~\ref{tab:allcomp} lists all \NComp{} comparisons with normal-approximation $p$-values from the bootstrap standard error, Holm-corrected jointly, because with 1,000 resamples the empirical $p$-values cannot fall below 0.002. The intervals describe test-set sampling for fixed trained models. An AUROC difference below 0.005 was fixed in advance as practically negligible.

Four analyses were added after the pre-specified ones and are reported as such: DenseNet121 trained with the EVA-X recipe (family C), all five ImageNet backbones trained with it (family E), seed replicates, and a rerun of ViT-B/16 at a lower learning rate after the pre-specified run diverged.

\section{Results}\label{sec:results}

\begin{table*}[t]
\caption{Test-Set Results for Single Models and Ensembles (25{,}596 Images, 2{,}797 Patients)}
\label{tab:models}
\centering
\footnotesize
\setlength{\tabcolsep}{4pt}
\begin{tabular}{llcccccr}
\toprule
Model & Recipe & AUROC & AUPRC & Macro-F1 & ECE & ms/image & Params (M) \\
\midrule
ResNet18 & Shared & 0.791 & 0.255 & 0.312 & 0.390 & 0.07 & 11 \\
ResNet18 & EVA-X & 0.794 & 0.257 & 0.322 & 0.020 & 0.07 & 11 \\
DenseNet121 & Shared & 0.808 & 0.284 & 0.339 & 0.390 & 0.40 & 7 \\
DenseNet121 & EVA-X & 0.805 & 0.260 & 0.320 & 0.014 & 0.40 & 7 \\
CoAtNet-0 & Shared & 0.800 & 0.269 & 0.328 & 0.386 & 0.35 & 27 \\
CoAtNet-0 & EVA-X & 0.800 & 0.255 & 0.313 & 0.022 & 0.35 & 27 \\
ViT-B/16$^\dagger$ & Shared & 0.705 & 0.153 & 0.211 & 0.441 & 0.28 & 86 \\
ViT-B/16, lr $3{\times}10^{-5}$ & Shared & 0.787 & 0.244 & 0.300 & 0.378 & 0.28 & 86 \\
ViT-B/16 & EVA-X & 0.785 & 0.236 & 0.291 & 0.019 & 0.28 & 86 \\
Swin-B & Shared & 0.792 & 0.260 & 0.324 & 0.376 & 0.71 & 87 \\
Swin-B & EVA-X & 0.802 & 0.260 & 0.318 & 0.024 & 0.71 & 87 \\
EVA-X-S & Shared & 0.798 & 0.263 & 0.318 & 0.378 & 0.23 & 22 \\
EVA-X-S & EVA-X & 0.828 & 0.309 & 0.360 & 0.012 & 0.23 & 22 \\
\midrule
Five ImageNet models, averaged & Shared & 0.814 & 0.283 & 0.337 & 0.397 & 1.81 & 217 \\
Five ImageNet models, label-wise stacking & Shared & 0.818 & 0.293 & 0.349 & 0.010 & 1.81 & 217 \\
Five ImageNet models, averaged & EVA-X & 0.817 & 0.283 & 0.336 & 0.013 & 1.81 & 217 \\
Five ImageNet + EVA-X-S, averaged & Mixed & 0.822 & 0.302 & 0.358 & 0.330 & 2.04 & 239 \\
Five ImageNet + EVA-X-S, averaged & EVA-X & 0.826 & 0.303 & 0.347 & 0.012 & 2.04 & 239 \\
DenseNet121 + EVA-X-S, averaged & EVA-X & 0.829 & 0.311 & 0.364 & 0.010 & 0.63 & 29 \\
\bottomrule
\end{tabular}
\\[2pt]{\footnotesize Macro-averaged over 14 findings. Parameters and latency are summed from unrounded per-model values. F1 thresholds and fitted fusers use validation data only. Latency: summed per-model bf16 inference at batch 64 on one A100 80GB. $^\dagger$Training diverged at epoch 8; the pre-specified run, kept in all ensembles. Shared: focal loss with positive weighting, Mixup, AdamW $10^{-4}$, batch 32. EVA-X: plain BCE, no Mixup, AdamW $2.5{\times}10^{-4}$, weight decay 0.05, batch 256, layer decay for EVA-X only.}
\end{table*}

\begin{table*}[t]
\caption{Paired Comparisons on the Test Set (Patient-Level Bootstrap, 1{,}000 Resamples)}
\label{tab:comparisons}
\centering
\footnotesize
\setlength{\tabcolsep}{4pt}
\begin{tabular}{lcccccc}
\toprule
Comparison (A vs.\ B) & Family & $\Delta$AUROC & 95\% CI & $p_{\text{Holm}}$ & $\Delta$AUPRC & $\Delta$Macro-F1 \\
\midrule
Five-model ImageNet ensemble vs.\ EVA-X-S & B & $-0.0136$ & $[-0.0178, -0.0098]$ & 0.012 & $-0.027$ & $-0.024$ \\
Ensemble of all six vs.\ EVA-X-S & B & $-0.0056$ & $[-0.0095, -0.0021]$ & 0.012 & $-0.007$ & $-0.002$ \\
DenseNet121 + EVA-X-S vs.\ EVA-X-S (same recipe)$^\ast$ & C & $+0.0013$ & $[-0.0003, +0.0031]$ & 0.228 & $+0.002$ & $+0.004$ \\
EVA-X-S vs.\ DenseNet121 (same recipe)$^\ast$ & C & $+0.0228$ & $[+0.0187, +0.0265]$ & 0.010 & $+0.050$ & $+0.041$ \\
Five-model ImageNet ensemble vs.\ DenseNet121 & A & $+0.0064$ & $[+0.0034, +0.0094]$ & 0.008 & $-0.001$ & $-0.003$ \\
Full stacking vs.\ calibrated averaging (ImageNet five) & A & $-0.0015$ & $[-0.0042, +0.0009]$ & 0.210 & $+0.003$ & $+0.003$ \\
Five-model ImageNet ensemble vs.\ EVA-X-S (shared recipe) & D & $+0.0166$ & $[+0.0118, +0.0208]$ & 0.014 & $+0.020$ & $+0.018$ \\
Five-model ImageNet ensemble vs.\ EVA-X-S (all EVA-X recipe)$^\ast$ & E & $-0.0109$ & $[-0.0142, -0.0076]$ & 0.014 & $-0.027$ & $-0.024$ \\
Ensemble of all six vs.\ EVA-X-S (all EVA-X recipe)$^\ast$ & E & $-0.0022$ & $[-0.0046, +0.0003]$ & 0.080 & $-0.006$ & $-0.014$ \\
\bottomrule
\end{tabular}
\\[2pt]{\footnotesize Positive values favor A. Ensembles use the fusion rule with the best out-of-fold validation AUROC, which was probability averaging in every family. $p$: two-sided bootstrap, $2(k+1)/(B+1)$ with $k$ the smaller tail count and $B=1{,}000$ (smallest attainable value 0.002), Holm-corrected within each family: A, ImageNet five (4 comparisons); B, with EVA-X-S under its own recipe (6); D, with EVA-X-S under the shared recipe (7), all pre-specified; C, the same-recipe pair (5), and E, all six models with the EVA-X recipe (7), both added after seeing family B (marked $^\ast$). Table~\ref{tab:allcomp} lists all 29 comparisons, with an approximate, non-confirmatory joint correction as a sensitivity analysis.}
\end{table*}

\subsection{Foundation Model versus Ensemble}
Table~\ref{tab:models} lists every model and ensemble, and Table~\ref{tab:comparisons} the paired comparisons. Fine-tuned with its own recipe, EVA-X-S reached \EvaxAUROC{} test AUROC. The five-model ImageNet ensemble (probability averaging, the rule selected on validation) reached \FiveAvgAUROC, lower by \CAD{} (95\% CI \CALo{} to \CAHi), and it also trailed on AUPRC (\CAPR) and macro-F1 (\CAF). The ensemble costs \FiveCost{} ms per image and \FiveParams{}M parameters, against \EvaxCost{} ms and \EvaxParams{}M for EVA-X-S.

Adding the five ImageNet models to EVA-X-S made it worse: the six-model average scored \SixAvgAUROC, \CBD{} (\CBLo{} to \CBHi) relative to EVA-X-S alone. This average was also badly calibrated (ECE \SixAvgECE) because the ImageNet members' class-weighted probabilities dominate. Averaging EVA-X-S with DenseNet121 trained with the same recipe changed AUROC by \CCD{} (\CCLo{} to \CCHi): the interval lies inside the pre-set $\pm$0.005 bound and includes zero, so for this pair the gain is negligible.

\subsection{Pretraining or Training Recipe?}
The foundation model's advantage depended on its recipe. Under the shared recipe used by the ImageNet backbones, EVA-X-S reached only \EvaxSharedAUROC, and the ImageNet ensemble outperformed it by \CGD{} (\CGLo{} to \CGHi). To test whether the recipe, rather than the pretraining, explained EVA-X-S's lead, we trained DenseNet121 with the EVA-X recipe. It scored \DenseEvaxRecAUROC, close to its \DenseAUROC{} under the shared recipe, and EVA-X-S exceeded it by \CDD{} (\CDLo{} to \CDHi) with the training settings held fixed. We then trained the other four backbones the same way (family E). Single-model AUROC changed by at most 0.01 (Table~\ref{tab:models}), apart from ViT-B/16, which diverged under the shared recipe and trained stably with this one (\VitEvaxRecAUROC, similar to its lower-learning-rate rerun). Their five-model average scored \FiveFullAUROC, against \FiveAvgAUROC{} under the shared recipe, and trailed EVA-X-S by \CHD{} (\CHLo{} to \CHHi). Averaging all six changed AUROC by \CID{} (\CILo{} to \CIHi) relative to EVA-X-S, inside the 0.005 bound. With the recipe held fixed for every model, the ensemble still does not reach the foundation model. The models still differ in architecture as well as pretraining, so these controls do not isolate pretraining. The foundation model, in contrast, is sensitive to how it is fine-tuned.

\textbf{Training variability.} Across seeds 42, 1, and 2, EVA-X-S scored \SeedEvaxMin{} to \SeedEvaxMax{} (s.d.\ \SeedEvaxSD) and DenseNet121 \SeedDenseMin{} to \SeedDenseMax{} (s.d.\ \SeedDenseSD). The EVA-X-S lead over DenseNet121 ranged from \SeedGapMin{} to \SeedGapMax{} by seed, and the shared-recipe five-model ensemble trailed each EVA-X-S seed by \SeedFiveGapMax{} to \SeedFiveGapMin. For these two models, seed-to-seed variation over three runs was an order of magnitude smaller than the headline gap. The other backbones, and therefore the ensembles, were trained once, so ensemble training variability was not measured.

\subsection{Fusion Rules and Calibration}
Among the ImageNet backbones, the ensemble improved on the best single model, DenseNet121, by \CED{} (\CELo{} to \CEHi) AUROC (Table~\ref{tab:fusion}). Full stacking did not outperform calibrated averaging in any of the six analyses (differences from \StackCalMinAll{} to \StackCalMaxAll), although it was the configuration of our original study. The large losses (below $-0.02$) occurred only when the members were trained with plain cross-entropy, whose probabilities are small (median about 0.01). They appear equally in out-of-fold validation, so they are not specific to the test set, and label-wise stacking shows the same pattern. This is consistent with L2-penalized logistic regression underfitting on small-scale probability features; we did not test stacking on standardized logits. Label-wise stacking scored nominally higher than calibrated averaging on test (\LWCalMin{} to \LWCalMax), but not on out-of-fold validation, where probability averaging was highest (Table~\ref{tab:fusion}); these differences were not tested. Every fitted rule reduced ECE from \FiveAvgECE{} for raw averaging to about 0.01, and calibrated averaging alone achieved \FiveCalAvgECE: the calibration benefit of stacking is recalibration. The miscalibration itself came from the class-weighted loss; DenseNet121 and EVA-X-S trained with plain cross-entropy had ECE near 0.01 (Table~\ref{tab:models}).

\begin{table}[t]
\caption{Fusion Rules for the Five ImageNet Backbones}
\label{tab:fusion}
\centering
\footnotesize
\setlength{\tabcolsep}{2pt}
\begin{tabular}{lcccccc}
\toprule
 & Val (OOF) & \multicolumn{5}{c}{Test} \\ \cmidrule(lr){2-2}\cmidrule(lr){3-7}
Method & AUROC & AUROC & AUPRC & Macro-F1 & ECE & Brier \\
\midrule
Prob.\ average & 0.8447 & 0.814 & 0.283 & 0.337 & 0.397 & 0.2210 \\
Calibrated avg. & 0.8400 & 0.813 & 0.283 & 0.342 & 0.013 & 0.0579 \\
Weighted avg. & 0.8436 & 0.817 & 0.292 & 0.349 & 0.011 & 0.0571 \\
Label-wise stack. & 0.8442 & 0.818 & 0.293 & 0.349 & 0.010 & 0.0570 \\
Full stacking & 0.8346 & 0.811 & 0.286 & 0.345 & 0.010 & 0.0572 \\
\bottomrule
\end{tabular}
\\[2pt]{\footnotesize All fusers, calibrators, and thresholds are fit on validation data. Val (OOF): patient-grouped 5-fold out-of-fold validation AUROC, the selection criterion; probability averaging scored highest and is the selected rule. Test differences between rules other than full stacking were not tested.}
\end{table}

\subsection{Ensemble Size and Cost}
\begin{figure}[t]
\centering
\includegraphics[width=0.85\columnwidth]{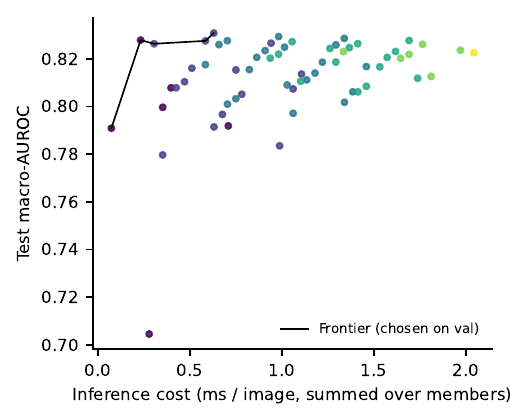}
\caption{Test macro-AUROC of every subset of the six models (shared-recipe ImageNet models and EVA-X-S with its own recipe, calibrated averaging) against summed backbone latency. The line connects subsets that improve validation AUROC as cost increases; color encodes the number of members.}
\label{fig:pareto}
\end{figure}

Fig.~\ref{fig:pareto} plots every subset of the six models against cost. This search uses the shared-recipe ImageNet models and calibrated averaging, so its values differ slightly from Table~\ref{tab:models}. Selected on validation, the best subset had \SubFMBestSize{} members (shared-recipe DenseNet121 and EVA-X-S, test AUROC \SubFMBestAUROC, against \SubFMA{} for EVA-X-S alone); larger subsets scored \SubFMC, \SubFMD, \SubFME, and \SubFMF{} for three to six members. This pair was selected among 63 subsets and has no paired test; the pre-specified pair test (Table~\ref{tab:comparisons}, family C) found no meaningful gain. Without the foundation model, the best ImageNet subset also had \SubFiveBestSize{} members (\SubFiveBestAUROC), against \SubFiveE{} for all five with calibrated averaging.

\subsection{Per-Finding Results}
\begin{figure}[t]
\centering
\includegraphics[width=0.85\columnwidth]{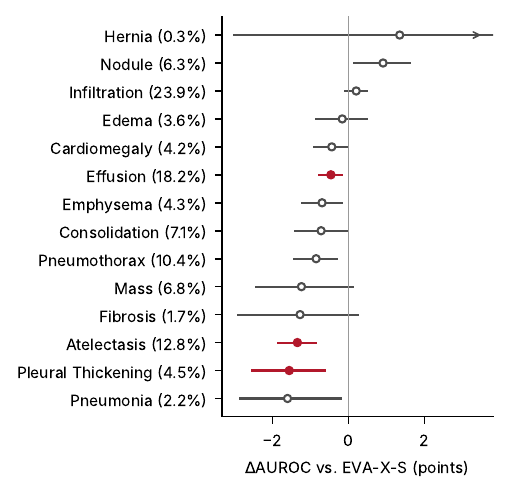}
\caption{Per-finding AUROC change of the six-model average relative to EVA-X-S, with 95\% patient-level bootstrap intervals; test prevalence in parentheses. Filled red: significant after Holm correction across the 14 findings. The Hernia interval is truncated (86 positives).}
\label{fig:perlabel}
\end{figure}

Relative to EVA-X-S, the six-model average was lower on \PLNeg{} of 14 findings (Fig.~\ref{fig:perlabel}). After Holm correction across findings, it was significantly worse for \PLSigNegNames, and significantly better for none. Per-finding results are descriptive; the primary endpoint is macro-AUROC.

\subsection{Training-Recipe Ablations}
For DenseNet121 under the shared recipe, removing augmentation changed test macro-F1 by \AblNoAug{} and removing the focal term by \AblNoFocal{} (Appendix). These sweeps report macro-F1, the metric of our original study. The focal parameter and learning rate were fixed before any test evaluation, and no setting was changed because of these results; the sweeps are descriptive.

\section{Discussion}\label{sec:discussion}
On this benchmark, a single small model pretrained on chest radiographs outperformed an ensemble of five ImageNet-pretrained networks that was ten times larger and nearly eight times slower in summed backbone latency, and combining them helped little: the small combinations selected on validation changed AUROC by at most 0.003, inside the pre-set negligible bound, and adding all five ImageNet models lowered it. Three observations qualify this result. First, the ImageNet ensemble did improve on its own best member, by \CED{} AUROC, consistent with the roughly 0.01 gains reported elsewhere~\cite{ashraf2023synth,li2024chexds,hsieh2026hybrid}; for these trained models the gain is distinguishable from zero but smaller than the gap to EVA-X-S, and it does not extend to AUPRC or macro-F1. Second, the recipe mattered asymmetrically. Switching EVA-X-S from the shared recipe to its own changed its AUROC by about 0.03, while the same switch moved the ImageNet ensemble only from \FiveAvgAUROC{} to \FiveFullAUROC. A comparison that trains every model with the ImageNet recipe would reach the opposite conclusion, so foundation models should be fine-tuned with settings suited to them. Third, our full-stacking configuration, the method our original study proposed, added nothing over calibrated averaging; other stacking designs, such as stacking on standardized logits, were not evaluated.

\textbf{Published results.} Our validation-selected EVA-X-S (\EvaxAUROC) is slightly below the published \PubEvax{}, which was obtained with test-set checkpoint selection and a different augmentation pipeline. The five-model ensemble (\FiveAvgAUROC) is comparable to published ImageNet-pretrained results on the official split, such as SwinCheX at 0.810~\cite{swinchex}.

\section{Limitations}
Bootstrap intervals reflect test-set sampling for fixed trained models. Seed replicates cover EVA-X-S and DenseNet121 only; the other ImageNet backbones, and therefore the ensembles, were trained once. Layer decay, defined over a Vision Transformer's block structure, was applied to EVA-X-S but not to the ImageNet backbones; it could also be applied to ViT-B/16, and we did not test whether doing so narrows the gap. EVA-X-S reached its best validation AUROC at epochs 25 to 30 of its 30-epoch limit and may be slightly undertrained. The pre-specified ViT-B/16 run diverged and remains in the shared-recipe ensembles. With the stable rerun (\VitFixAUROC{} instead of \VitAUROC) the five-model average scored \VitFixFiveAUROC, \VitFixFiveGap{} relative to EVA-X-S, a point estimate without a paired interval; under the EVA-X recipe (family E), where ViT-B/16 trained stably, the corresponding paired difference is \CHD{} (\CHLo{} to \CHHi). EVA-X's pretraining used unlabeled images from ChestX-ray14, CheXpert, and MIMIC-CXR; its authors report excluding test images, which we could not verify independently, and part of its advantage is additional in-domain data. We used ImageNet input normalization for EVA-X-S rather than the statistics in its code, a departure from its recipe whose effect we did not measure. We evaluated one foundation model; others, such as RAD-DINO or vision-language models, may behave differently. The weighted-average fuser optimizes log-loss from equal starting weights and can miss better combinations; it was not the selected rule. Validation AUROC exceeded test AUROC by 0.02 to 0.03 for every system; one possible contributor is that the official test set has more findings per abnormal image than the training and validation sets (Table~\ref{tab:splits}), but we did not establish the cause. The paired comparisons estimate differences between systems on the same test patients and apply to this test population. All results come from one dataset with report-mined labels~\cite{oakden2020}, at $224\times224$ resolution, where pretrained-model rankings can differ from those at higher resolution~\cite{tayebi2025dinov3}, without external validation. Subgroup results cover the ensembles only.

\section{Ethical Considerations}
This study uses only the public, de-identified NIH ChestX-ray14 dataset. The models are research prototypes, not intended for clinical use, which would require external and prospective validation, regulatory review, and monitoring. Subgroup results (Appendix) cover only sex and age, the attributes the dataset records.

\section{Conclusion}
A chest X-ray foundation model with 22M parameters, fine-tuned with its own recipe, outperformed a five-backbone ImageNet ensemble on ChestX-ray14 at about one-eighth of the summed backbone latency; the validation-selected pairs with one ImageNet model changed AUROC by at most 0.003 and adding all five lowered it, and the advantage held when every ImageNet backbone was trained the same way and across three EVA-X-S seeds against the fixed ensemble, which was itself trained once. On ChestX-ray14 at $224\times224$, a domain-pretrained model was a better use of compute than architectural diversity, provided it was fine-tuned appropriately, and our full-stacking configuration delivered no more than calibrated averaging.

\section*{Code and Data Availability}
Code for data preparation, training, verification, analysis, and every table and figure is available at \url{https://github.com/adityaseelaboyina/cxr-ensemble-vs-foundation-model}. ChestX-ray14 is publicly available from the NIH Clinical Center.

\appendix
\begin{table}[t]
\caption{Dataset Splits (Avg.\ Findings: Mean Number of Findings per Abnormal Image)}
\label{tab:splits}
\centering
\begin{tabular}{lrrc}
\toprule
Split & Images & Patients & Avg.\ Findings \\
\midrule
Train & 77,988 & 25,207 & 1.50 \\
Validation & 8,536 & 2,801 & 1.49 \\
Test & 25,596 & 2,797 & 1.73 \\
\bottomrule
\end{tabular}
\end{table}

\begin{table}[t]
\caption{Training Ablations on DenseNet121 (Test Macro-F1)}
\label{tab:ablation_single}
\centering
\begin{tabular}{lcc}
\toprule
Configuration & Macro-F1 & $\Delta$ \\
\midrule
Full training recipe & 0.339 & --- \\
\quad No focal loss & 0.332 & $-0.007$ \\
\quad No augmentation & 0.314 & $-0.025$ \\
\bottomrule
\end{tabular}
\end{table}

\begin{table}[t]
\caption{Focal Loss $\gamma$ Sensitivity (DenseNet121, Test Macro-F1)}
\label{tab:gamma}
\centering
\begin{tabular}{ccc}
\toprule
$\gamma$ & Macro-F1 & $\Delta$ \\
\midrule
0 (BCE) & 0.332 & --- \\
1 & 0.326 & $-0.006$ \\
2 & 0.339 & $+0.007$ \\
3 & 0.329 & $-0.003$ \\
\bottomrule
\end{tabular}
\end{table}

\begin{table}[t]
\caption{Initial Learning Rate Sensitivity (DenseNet121, Test Macro-F1)}
\label{tab:lr}
\centering
\begin{tabular}{cc}
\toprule
Learning Rate & Macro-F1 \\
\midrule
$10^{-3}$ & 0.223 \\
$10^{-4}$ & 0.339 \\
$5\times10^{-5}$ & 0.331 \\
\bottomrule
\end{tabular}
\end{table}

\begin{table}[t]
\caption{Test Macro-AUROC by Patient Subgroup}
\label{tab:subgroups}
\centering
\footnotesize
\begin{tabular}{lrrcc}
\toprule
Subgroup & Images & Patients & Five & Six \\
\midrule
Male & 14,882 & 1,557 & 0.813 & 0.821 \\
Female & 10,714 & 1,240 & 0.814 & 0.822 \\
Age $<$40 & 8,290 & 1,048 & 0.757 & 0.764 \\
Age 40--59 & 11,224 & 1,169 & 0.810 & 0.818 \\
Age $\geq$60 & 6,078 & 686 & 0.806 & 0.814 \\
\bottomrule
\end{tabular}
\\[2pt]{\footnotesize Probability averaging, the selected rule. Five: the five ImageNet backbones. Six: the five ImageNet backbones plus EVA-X-S (own recipe). Age is recorded per image, so a patient imaged at different ages can appear in more than one age group. Some rare findings have few positives within a subgroup, so subgroup estimates are imprecise and no intervals are shown.}
\end{table}

\begin{table}[t]
\caption{Per-Finding Test Results of the Five-Model ImageNet Ensemble (Full Stacking)}
\label{tab:perclass}
\centering
\footnotesize
\setlength{\tabcolsep}{4pt}
\begin{tabular}{lccccr}
\toprule
Finding & Precision & Recall & F1 & AUC & Support \\
\midrule
Atelectasis & 0.330 & 0.458 & 0.384 & 0.763 & 3279 \\
Cardiomegaly & 0.399 & 0.432 & 0.415 & 0.899 & 1069 \\
Effusion & 0.441 & 0.684 & 0.536 & 0.829 & 4658 \\
Infiltration & 0.355 & 0.762 & 0.484 & 0.710 & 6112 \\
Mass & 0.343 & 0.392 & 0.366 & 0.818 & 1748 \\
Nodule & 0.230 & 0.392 & 0.289 & 0.759 & 1623 \\
Pneumonia & 0.050 & 0.389 & 0.088 & 0.704 & 555 \\
Pneumothorax & 0.410 & 0.607 & 0.490 & 0.864 & 2665 \\
Consolidation & 0.157 & 0.430 & 0.230 & 0.737 & 1815 \\
Edema & 0.202 & 0.352 & 0.257 & 0.853 & 925 \\
Emphysema & 0.275 & 0.724 & 0.398 & 0.916 & 1093 \\
Fibrosis & 0.156 & 0.145 & 0.150 & 0.818 & 435 \\
Pleural Thickening & 0.159 & 0.336 & 0.216 & 0.773 & 1143 \\
Hernia & 0.577 & 0.477 & 0.522 & 0.912 & 86 \\
\bottomrule
\end{tabular}
\end{table}

\begin{table*}[t]
\caption{All 29 Paired AUROC Comparisons}
\label{tab:allcomp}
\centering
\footnotesize
\setlength{\tabcolsep}{4pt}
\begin{tabular}{llcccccc}
\toprule
 & & & & \multicolumn{2}{c}{Primary (empirical)} & \multicolumn{2}{c}{Sensitivity (normal)} \\ \cmidrule(lr){5-6}\cmidrule(lr){7-8}
Fam. & A vs.\ B & $\Delta$ & 95\% CI & $p$ & $p_{\text{fam}}$ & $p_{N}$ & $p_{N,\text{all}}$ \\
\midrule
A & five average vs.\ DenseNet121 & $+0.0064$ & $[+0.0034, +0.0094]$ & 0.002 & 0.008 & $<$0.001 & $<$0.001 \\
A & five full stacking vs.\ five calibrated avg. & $-0.0015$ & $[-0.0042, +0.0009]$ & 0.210 & 0.210 & 0.234 & 0.468 \\
A & five full stacking vs.\ five average & $-0.0032$ & $[-0.0057, -0.0008]$ & 0.012 & 0.024 & 0.013 & 0.091 \\
A & five calibrated avg. vs.\ DenseNet121 & $+0.0048$ & $[+0.0018, +0.0080]$ & 0.002 & 0.008 & 0.003 & 0.025 \\
B & six average vs.\ EVA-X-S & $-0.0056$ & $[-0.0095, -0.0021]$ & 0.004 & 0.012 & 0.003 & 0.025 \\
B & six full stacking vs.\ six calibrated avg. & $-0.0045$ & $[-0.0071, -0.0021]$ & 0.002 & 0.012 & $<$0.001 & 0.004 \\
B & six full stacking vs.\ six average & $-0.0042$ & $[-0.0068, -0.0017]$ & 0.004 & 0.012 & 0.001 & 0.012 \\
B & six calibrated avg. vs.\ EVA-X-S & $-0.0053$ & $[-0.0084, -0.0022]$ & 0.002 & 0.012 & $<$0.001 & 0.009 \\
B & five average vs.\ EVA-X-S & $-0.0136$ & $[-0.0178, -0.0098]$ & 0.002 & 0.012 & $<$0.001 & $<$0.001 \\
B & six average vs.\ five average & $+0.0080$ & $[+0.0074, +0.0086]$ & 0.002 & 0.012 & $<$0.001 & $<$0.001 \\
D & six average vs.\ DenseNet121 & $+0.0091$ & $[+0.0060, +0.0126]$ & 0.002 & 0.014 & $<$0.001 & $<$0.001 \\
D & six full stacking vs.\ six calibrated avg. & $-0.0029$ & $[-0.0054, -0.0006]$ & 0.014 & 0.014 & 0.020 & 0.119 \\
D & six full stacking vs.\ six average & $-0.0042$ & $[-0.0067, -0.0019]$ & 0.002 & 0.014 & $<$0.001 & 0.007 \\
D & six calibrated avg. vs.\ DenseNet121 & $+0.0078$ & $[+0.0047, +0.0112]$ & 0.002 & 0.014 & $<$0.001 & $<$0.001 \\
D & five average vs.\ EVA-X-S & $+0.0166$ & $[+0.0118, +0.0208]$ & 0.002 & 0.014 & $<$0.001 & $<$0.001 \\
D & six average vs.\ EVA-X-S & $+0.0193$ & $[+0.0151, +0.0229]$ & 0.002 & 0.014 & $<$0.001 & $<$0.001 \\
D & six average vs.\ five average & $+0.0027$ & $[+0.0020, +0.0035]$ & 0.002 & 0.014 & $<$0.001 & $<$0.001 \\
C & pair average vs.\ EVA-X-S & $+0.0013$ & $[-0.0003, +0.0031]$ & 0.114 & 0.228 & 0.132 & 0.397 \\
C & pair full stacking vs.\ pair calibrated avg. & $-0.0324$ & $[-0.0383, -0.0271]$ & 0.002 & 0.010 & $<$0.001 & $<$0.001 \\
C & pair full stacking vs.\ pair average & $-0.0328$ & $[-0.0387, -0.0274]$ & 0.002 & 0.010 & $<$0.001 & $<$0.001 \\
C & pair calibrated avg. vs.\ EVA-X-S & $+0.0010$ & $[-0.0007, +0.0028]$ & 0.266 & 0.266 & 0.284 & 0.468 \\
C & EVA-X-S vs.\ DenseNet121 & $+0.0228$ & $[+0.0187, +0.0265]$ & 0.002 & 0.010 & $<$0.001 & $<$0.001 \\
E & six average vs.\ EVA-X-S & $-0.0022$ & $[-0.0046, +0.0003]$ & 0.080 & 0.080 & 0.073 & 0.294 \\
E & six full stacking vs.\ six calibrated avg. & $-0.0250$ & $[-0.0300, -0.0206]$ & 0.002 & 0.014 & $<$0.001 & $<$0.001 \\
E & six full stacking vs.\ six average & $-0.0256$ & $[-0.0304, -0.0213]$ & 0.002 & 0.014 & $<$0.001 & $<$0.001 \\
E & six calibrated avg. vs.\ EVA-X-S & $-0.0027$ & $[-0.0053, -0.0003]$ & 0.040 & 0.080 & 0.031 & 0.153 \\
E & five average vs.\ EVA-X-S & $-0.0109$ & $[-0.0142, -0.0076]$ & 0.002 & 0.014 & $<$0.001 & $<$0.001 \\
E & six average vs.\ five average & $+0.0087$ & $[+0.0074, +0.0101]$ & 0.002 & 0.014 & $<$0.001 & $<$0.001 \\
E & EVA-X-S vs.\ DenseNet121 & $+0.0228$ & $[+0.0187, +0.0265]$ & 0.002 & 0.014 & $<$0.001 & $<$0.001 \\
\bottomrule
\end{tabular}
\\[2pt]{\scriptsize Primary analysis: empirical bootstrap $p=2(k+1)/(B+1)$ and Holm within family ($p_{\text{fam}}$), as in Table~\ref{tab:comparisons}. Approximate sensitivity analysis only, not confirmatory: $p_N=2\{1-\Phi(|\Delta|/\mathrm{SE})\}$ with SE the bootstrap standard error; its finite-sample validity for patient-clustered macro-AUROC differences with rare findings is not established, and $p_{N,\text{all}}$, Holm across all 29. It is reported because with $B=1{,}000$ the empirical $p$ cannot fall below 0.002, so a joint Holm correction over 29 tests could not reach 0.05. Systems: five = the five ImageNet backbones; six = five plus EVA-X-S; pair = DenseNet121 and EVA-X-S. In C and E every model uses the EVA-X recipe. A: ImageNet five; B: EVA-X-S own recipe; C: same-recipe pair; D: EVA-X-S shared recipe; E: all six with the EVA-X recipe. C and E were added post hoc.}
\end{table*}

\end{document}